\documentclass[prd,twocolumn,showpacs,preprintnumbers,superscriptaddress,floatfix,nofootinbib]{revtex4-1}
\usepackage{multirow}
\usepackage{amsfonts}
\usepackage{graphicx,,booktabs}
\usepackage{amssymb}
\usepackage{amsmath,txfonts,ulem}
\usepackage{graphicx}
\usepackage{dcolumn}
\usepackage{color}
\usepackage{bm}
\usepackage{ulem}
\usepackage[colorlinks,
            citecolor=blue,
            anchorcolor=green,
            menucolor=orange,
            linkcolor=red,
            filecolor=red,
            runcolor=pink,
            urlcolor=blue,
            frenchlinks=red]{hyperref}

\allowdisplaybreaks[4]

\begin{document}

\title{\boldmath Analysis of the contribution of the quantum anomaly energy to the proton mass}

\author{Xiao-Yun Wang}
\email{Corresponding author: xywang@lut.edu.cn}
\affiliation{Department of physics, Lanzhou University of Technology,
Lanzhou 730050, China}
\affiliation{Lanzhou Center for Theoretical Physics, Key Laboratory of Theoretical Physics of Gansu Province, Lanzhou University, Lanzhou, Gansu 730000, China}

\author{Jingxuan Bu}
\email{jingxuanbu@yeah.net}
\affiliation{Department of physics, Lanzhou University of Technology,
Lanzhou 730050, China}

\author{Fancong Zeng}
\email{zengfc@ihep.ac.cn}
\affiliation{  Institute of High Energy Physics, Chinese Academy of Sciences, Beijing 100049, China}
\affiliation{ University of Chinese Academy of Sciences, Beijing 100049, China}

\date{\today}
\begin{abstract}
Inspired by the recent Hall C and GlueX measurements of $J/\psi$ photoproduction, a systematic analysis of the contribution of quantum anomalous energy (QAE) to the proton mass is carried out under the framework of the  vector meson dominance model.  The results show that the effective Pomeron model and the parametrized two gluon exchange model can explain the cross section of $J/\psi$ photoproduction well. Based on the predicted cross section values given by the two models, the distribution of the QAE contribution with the energy is extracted for the first time. Finally, the average value of the QAE contribution is estimated to be (3.50$\pm$0.70)$\%$, which suggests that the QAE contribution to the proton mass is small. Accordingly, we compared this result with those of other groups and explored the causes for the differences.

\end{abstract}


\maketitle

\section{introduction}

Nowadays, the research on the source of nucleon mass is an important topic in hadron physics, which helps us better understand the structure and dynamics of the nucleon system. According to the quantum chromodynamics (QCD) theory \cite{DJGF Un,DJGF As,DJGF ASYM}, nucleons are usually composed of quarks and massless gluons, however quarks make up only a small fraction of the mass of nucleons, and other sources of mass for nucleons are of great interest to researchers \cite{clorce,yh,am,ji94,ji95}. Among them, Ji obtained a four-part decomposition of the nucleon mass based on the study of the structure of the QCD energy-momentum tensor (EMT) and its matrix elements in nucleon states \cite{ji94,ji95}.

The derivation starts from the QCD EMT. The EMT can be split into traceless and trace parts which can be written as \cite{ji95}
\begin{align}\label{eq:EMT}
{T}^{\mu \nu}=\hat{T}^{\mu \nu}+\bar{T}^{\mu \nu}
\end{align}
here the composite operators can be    denoted as
\begin{equation}\label{eq:dEMT}
\begin{aligned}
&\hat{T}^{\mu \nu}=\hat{T}^{\mu \nu}_{a}(\mu^2)+\hat{T}^{\mu \nu}_{m}(\mu^2)
\\
&\bar{T}^{\mu \nu}=\bar{T}^{\mu \nu}_{q}(\mu^2)+\bar{T}^{\mu \nu}_{g}(\mu^2)
\end{aligned}
\end{equation}
In hadron state, the forward matrix element of EMT takes the form as \cite{ji95}
\begin{align}\label{eq:PTP}
\left< P|T^{\mu \nu}|P \right>=\frac{P^{\mu}P^{\nu}}{M}
\end{align}
  Reference \cite{ji95} assumes that the state is normalized as $\left< P|P \right>=(E^2/M)(2\pi)^3\delta^3(0)$, where $E$ and $M$ are the energy  and mass of  the hadron, respectively. Using the Lorentz symmetry, both sides of Eq.  \ref{eq:PTP}  can be divided into trace and traceless parts. The corresponding individual forward matrix elements can be written as   \cite{ji95}
\begin{equation}\label{eq:PTPfour}
\begin{aligned}
&\left< P|\hat{T}^{\mu \nu}_{m}|P \right>=b(\mu^2)\frac{1}{4}g^{\mu \nu}M
\\
&\left< P|\hat{T}^{\mu \nu}_{a}|P \right>=(1-b(\mu^2))\frac{1}{4}g^{\mu \nu}M
\\
&\left< P|\bar{T}^{\mu \nu}_{q}|P \right>=a(\mu^2)(P^{\mu}P^{\nu}-\frac{1}{4}g^{\mu \nu}M^2)/M
\\
&\left< P|\bar{T}^{\mu \nu}_{g}|P \right>=(1-a(\mu^2))(P^{\mu}P^{\nu}-\frac{1}{4}g^{\mu \nu}M^2)/M
\end{aligned}
\end{equation}
These four parts of the EMT are determined by the momentum fraction $a(\mu^2)$ and trace anomaly parameter $b(\mu^2)$ \cite{ji95,wr}.
Based on the above, the breakdown for the Hamiltonian is given as \cite{ji95}

\begin{align}\label{eq:Hsum}
H_{QCD}=H_{q}+H_{g}+H_{m}+H_{a}
\end{align}
where $H_{q}$ and $H_{g}$ represent the total energy of the quarks and  gluons, respectively; $H_{m}$ is the quark mass contribution; $H_{a}$ is the anomaly contribution. The hadron mass is obtained by calculating the expectation of the Hamiltonian operator in the rest frame of hadrons \cite{ji95}
\begin{align}\label{eq:Mn}
M=\frac{\left< P|H_{QCD}|P \right>}{\left< P|P \right>}\Bigg|_{\rm rest\ frame}
\end{align}
According to Eqs.  (\ref{eq:Hsum}) and (\ref{eq:Mn}),   the four parts of the hadron mass can be written as
\begin{equation}\label{eq:Mnfour}
\begin{aligned}
&M_{q}=\frac{3}{4}\ (a-\frac{b}{1+\gamma_{m}})M
\\
&M_{g}=\frac{3}{4}\ (1-a)M
\\
&M_{m}=\frac{4+\gamma_{m}}{4(1+\gamma_{m})}bM
\\
&M_{a}=\frac{1}{4}(1-b)M
\end{aligned}
\end{equation}where $\gamma_{m}$ is the anomalous dimension of the quark mass \cite{gama}.
 The last part of Eq. (\ref{eq:Mnfour}) is the quantum anomalous energy (QAE).  The QAE is a new source exists in the quantum field theories,  which has been widely studied in recent years as a key to understanding the origin of the proton mass.    Recently Ref. \cite{decompo} argues that it arises from the breaking of scale symmetry caused by UV divergences in the quantum field theories.

We note that some work discussed other ways to determine the contribution of QAE. For example, one work linked the QAE contribution  with the $J/\psi$ near-threshold photoproduction from the method of the  holographic calculation \cite{Hatta:2019lxo,Hatta:2018ina,Hatta:2018sqd}. What is more,  Ref. \cite{new data} discussed  the differential cross section with
the maximal and minimal trace anomaly contribution to the
EMT matrix element \cite{Hatta:2019lxo,Hatta:2018ina,Hatta:2018sqd}. However, the fitting results show that the curves corresponding to the maximum and minimum trace anomaly contributions are very close, which makes it difficult to determine the contribution of QAE.
In fact, the contribution of QAE can also be extracted from the cross section of vector mesons photoproduction at the threshold under the framework of the  vector meson dominance (VMD) model \cite{vmd,wr}. Fortunately, the GlueX and Hall C collaborations in Jefferson Lab have measured some new $J/\psi$ photoproduction data \cite{glue,new data}, which provides a very good opportunity for us to study the magnitude of the QAE contribution. In fact, theoretical and experimental researchers have extracted the QAE contributions based on Hall C and GlueX data. However, we note that the QAE values obtained from these extractions have a strong energy dependence, showing that the QAE values extracted from the cross sections at higher energies are also larger \cite{new data},  which poses substantial difficulties to the final determination of the size of the QAE contribution. Therefore, it is necessary to give an overall picture of how the QAE varies with the energy distribution at the threshold. Since the energy interval of the current experimental data is large, and the measured minimum energy point is several hundred MeV away from the threshold energy \cite{glue,new data}, we need to introduce relevant theoretical models to first fit the cross section of $J/\psi$ photoproduction. Afterwards, combined with the predicted value of the model and the experimental data, the calculation of the QAE distribution with the energy can be carried out.

In our calculations, the parametrized two gluon exchange model and the effective Pomeron model will be employed to fit the cross section of $J/\psi$ photoproduction. The two gluon exchange model proposed in Ref. \cite{two1} is directly related to the gluon distribution function. Due to the current gluon distribution functions such as the function from  GRV98 \cite{14}, NNPDF \cite{15}, CJ15 \cite{16,17}, and IMParton16 \cite{18} cannot interpret well the vector meson photoproduction data at near threshold; in this work the gluon distribution function is taken as $xg(x,m_{\psi}^2)=A_{0}x^{A1} (1-x)^{A2}$ \cite{xg1}. Here, $A_{0},\ A_{1},\ A_{2}$ are free parameters, which were determined in our previous work \cite{two2} by fitting the GlueX experimental data. Therefore, we proposed the parametrized two gluon exchange model, which can well predict the differential and total cross section of $J/\psi$ photoproduction in a wide energy range. And from Ref.\ \cite{po1}, Joint Physics Analysis Center (JPAC) collaboration proposed the effective Pomeron model that can effectively explain the $J/\psi$ photoproduction. In this work, we will systematically study the contribution of QAE based on the results given by these two models.

This paper is organized as follows. After the introduction, the employed models are presented in Sec. \ref{sec:model}. The analysis of the numerical results is in Sec. \ref{sec:result}, followed by a short summary in Sec. \ref{sec:sum}.

\section{FORMALISM}\label{sec:model}

\subsection{Parametrized two gluon exchange model}\label{tgm}

\begin{figure}[htbp]
	\centering
	\includegraphics[width=0.487\textwidth]{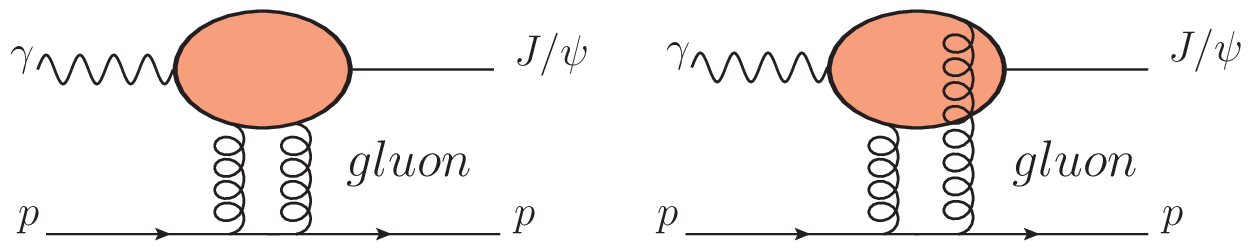}
\caption{The Feynman diagrams of the parametrized two gluon exchange model for $J/\psi$ photoproduction.}
  \label{Fig:two}
\end{figure}

The parametrized two gluon exchange model can be described in the exclusive vector meson photoproduction process, the photon fluctuation into the quark-antiquark pair $(\gamma \to q+\bar{q})$, and the quark-antiquark pair can be treated as a dipole, finally the dipole hadronization into the vector meson. The dipole interacts with a nucleon through two gluon exchange. The picture of the two gluon exchange process is shown in Fig. \ref{Fig:two}. The process is described as $\gamma N\to J/\psi N$.

In lowest order perturbative QCD, the exclusive vector meson photoproduction amplitude takes the form \cite{two1,f2,xg,two birth}

\begin{equation}\label{eq:tg amli'}
\begin{aligned}
\mathcal{T}=&\frac{i2\sqrt{2}\pi^2}{3}m_{q}\alpha_{s}e_{q}f_{V}F_{2g}(t)
\\
&\left[\frac{xg(x,Q_{0}^2)}{m_{q}^4}+\int_{Q_{0}^2}^{+\infty}\frac{dl^2}{m_{q}^2(m_{q}^2+l^2)}\frac{\partial xg(x,l^2)}{\partial l^2}\right]
\end{aligned}
\end{equation}
When the amplitude is normalized, there is $d\sigma/dt=\alpha |\mathcal{T}|^2$. The $J/\psi$ differential cross section is denoted as \cite{two1,two2}
\begin{align}\label{eq:do/dt}
\frac{d\sigma}{dt}=\frac{\pi^3\Gamma_{e^+ e^-}\alpha_{s}}{6\alpha m_{q}^5}[xg(x,m_{J/\psi}^2)]^2 exp(bt)
\end{align}
where the gluon distribution function takes the form as $xg(x,m_{J/\psi}^2)=A_{0}x^{A1} (1-x)^{A2}$  with   free parameters $A_{0},\ A_{1},\ A_{2}$  \cite{two2}. $W$ represents the c.m. energy of the $\gamma N\to J/\psi N $ reaction. The exponential part in the above formula can use the standard form \cite{two2}.
 The radiative decay $\Gamma_{e^+ e^-}$ is equal to 5.53 keV \cite{zlbj}. $\alpha_{s}$=0.5 is the QCD coupling constant  \cite{as}.
   The total cross section is obtained by integrating the differential cross section from $t_{\min}$ to $t_{\max}$ over the allowed kinematical range that
\begin{align}\label{eq:tmin tmax}
\sigma=\int_{t_{\min}}^{t_{\max}}\left(\frac{d\sigma}{dt}\right)dt
\end{align}
where the four-momentum transfers $t_{\min}$ and $t_{\max}$ are
\begin{align}\label{eq:tmax tmin}
t_{\max}(t_{\min})=\frac{m_{J/\psi}^4}{4W^{2}}-(p_{\gamma}\mp p_{J/\psi})^2
\end{align}
  The c.m. energies and momenta of the photon and produced $J/\psi$ meson can be written as

\begin{equation*}\label{eq:E}
\begin{aligned}
&E_{\gamma}=\frac{W^2-M_{N}^2}{2W},E_{J/\psi}=\frac{W^2+m_{J/\psi}^2-M_{N}^2}{2W}, \\
 & p_{\gamma}=E_{\gamma},p_{J/\psi}=\sqrt{E_{J/\psi}^2-m_{J/\psi}^2}.
\end{aligned}
\end{equation*}

\subsection{Effective Pomeron model}\label{pm}

\begin{figure}[htbp]
	\centering
	\includegraphics[width=0.28\textwidth]{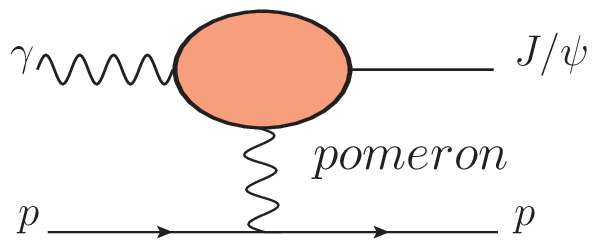}
\caption{The Feynman diagram of the effective Pomeron model for $J/\psi$ photoproduction.}
\label{Fig:pop}
\end{figure}

In general, the photoproduction of vector mesons at high energies is well described by Pomeron exchange. In Ref. \cite{po2}, it was shown that the Pomeron can be considered as an explicit two-gluon or three-gluon exchange. Based on these situations, the JPAC collaboration has constructed two effective Pomeron models for the photoproduction of vector mesons at low and high energies \cite{po1,poh}. In this work, we use the low-energy effective Pomeron model proposed by JPAC group \cite{po1} to describe the photoproduction cross section at $J/\psi$ near the threshold.

In the effective Pomeron model,  the photon fluctuation into the quark-antiquark pair, and finally the dipole converts into a vector meson. The interaction between the quark-antiquark pair and the nucleon proceeds via Pomeron exchange. The picture of the effective  Pomeron model is described in Fig. \ref{Fig:pop}.
The differential cross section of the $J/\psi$ photoproduction reaction  is given as \cite{po1}
\begin{align}\label{eq:p do}
\frac{d\sigma}{dt}=\frac{4\pi\alpha}{64\pi W^2 p_{\gamma}^2}\sum\limits_{\lambda_{\gamma},\lambda_{p},\lambda_{J/\psi},\lambda_{p^{\prime}}}\frac{1}{4}\left|\left\langle\lambda_{J/\psi}\lambda_{p^{\prime}}|T|\lambda_{\gamma}\lambda_{p}\right\rangle\right|^2
\end{align}
The total cross section can be calculated  by integrating the differential cross section. The covariant amplitude in Eq.  (\ref{eq:p do}) takes the form \cite{po1}
\begin{equation}\label{eq:p do am}
\begin{aligned}
\left\langle\lambda_{J/\psi}\lambda_{p^{\prime}}|T_{P}|\lambda_{\gamma}\lambda_{p}\right\rangle=&F(W,t)\bar{u}\left(p_{f},\lambda_{p^{\prime}}\right)\gamma_{\mu}u\left(p_{i},\lambda_{p}\right)
\\
&\times\left[\varepsilon^{\mu}\left(p_{\gamma},\lambda_{\gamma}\right)q^{v}-\varepsilon^{v}\left(p_{\gamma},\lambda_{\gamma}\right)q^{\mu}\right]
\\
&\times\varepsilon_{v}^{*}\left(p_{J/\psi},\lambda_{J/\psi}\right)
\end{aligned}
\end{equation}
Here, $u(p_{i},\lambda_{p})$ and $u(p_{f},\lambda_{p^{\prime}})$ are the Dirac spinors for the target and recoil protons. $q$, $p$, and $p^{\prime}$ are the photon,  the initial and final nucleon  four-momenta.   $\lambda_{p}$ and $\lambda_{p^{\prime}}$ are their helicities, respectively. $\varepsilon$ determine polarization of the photon and the $J/\psi$. Besides, the amplitude takes the form
\begin{align}\label{eq:p am f}
F(W,t)=iA_{J/\psi}\left(\frac{W^2-W_{thr}^2}{W_{0}^2}\right)^{{\alpha(t)}}\frac{e^{b_{0}(t-t_{\min})}}{W^2}
\end{align}
where $\alpha(t)=\alpha_{0}+\alpha^{\prime}t$ is the standard Pomeron trajectory given in Ref. \cite{amti}. Usually, the $\alpha_{0}$ and $\alpha^{\prime}$ are determined by fitting the vector meson photoproduction data at high or low energy. The parameters in the Pomeron trajectory are different because of the difference in the physical mechanism of the photoproduction of vector mesons at high energy or threshold. In this work, the free parameters $A_{J/\psi},\ \alpha_{0},\ \alpha^{\prime},\ b_{0}$ are obtained by fitting the low energy $J/\psi$ photoproduction data from GlueX \cite{glue} and SLAC \cite{slac} collaboration, as listed  in Table  \ref{apsi}. The energy scale parameter is given as $W_{0}$=1 GeV and $W_{thr} = M_{N} + m_{J/\psi}$ is the threshold energy.

\subsection{VMD model and the contribution of  QAE }\label{vmd}
At the energy threshold,
  the forward differential cross section of $\gamma N\to J/\psi N$ reaction under the framework of the VMD model takes the form as \cite{vmd}

\begin{align}\label{eq:b vmd}
\frac{d\sigma_{\gamma N\to J/\psi N}}{dt}\Bigg|_{t=t_{min}}=\frac{3  R^2~\Gamma_{e^+ e^-} }{\alpha m_{J/\psi}}\frac{d\sigma_{J/\psi N\to J/\psi N}}{dt}\Bigg|_{t=t_{min}}
\end{align}
where $R$ is the ratio between the final momenta $p_{J/\psi}$ and the initial momenta $p_{\gamma}$.
 The final part is the differential cross section of the $J/\psi -N$ interaction, which can be  given as
\begin{align}\label{eq:xn xn vmd}
\frac{d\sigma_{J/\psi N\to J/\psi N}}{dt}\Bigg|_{t=t_{min}}=\frac{1}{64}\frac{1}{m_{J/\psi}^2(\lambda^2-M_{N}^2)}|F_{J/\psi N}|^2
\end{align}
where $\lambda=(W^{2}-m^{2}_{J/\psi}-M^{2}_{N})/(2m_{J/\psi})$ is the nucleon energy \cite{vmd}, $F_{J/\psi N}$ is the $J/\psi-N$ elastic scattering amplitude, which can be written as  \cite{Rb,wr}
\begin{equation}\label{eq:F vmd}
\begin{aligned}
F_{J/\psi N}&\simeq r_{0}^3 d_{2}\frac{2\pi^2}{27}\left(2M_{N}^2-\left\langle N\left|\sum\limits_{h=u,d,s}m_{h}\bar q_{h}q_{h}\right|N\right\rangle\right)
\\
&=r_{0}^3 d_{2}\frac{2\pi^2}{27}2M_{N}^2(1-b)
\end{aligned}
\end{equation}
where the Bohr radius $r_{0}= \frac{4}{3\alpha_{s} m_{q} }  $ of the  $J/\psi$ meson is taken from Ref. \cite{Rb}. The parameter  $b$
is an important physical quantity in this work,   representing the QCD trace anomaly parameter away from the chiral limit.  In this paper, we set the  mass of the constituent quark (charm quark) $m_{q}$=1.67 GeV.
The Wilson coefficient $d_2$  in Eq.  (\ref{eq:F vmd}) can be described as \cite{Rb1}
\begin{align}
d_{n}  =\left(\frac{32}{N_{c}}\right)^2\sqrt{\pi}\frac{\Gamma(n+5/2)}{\Gamma(n+5)}
\end{align}

  Actually, the differential cross section in Eq.  (\ref{eq:b vmd}) at the  c.m. energy threshold can be written  as \cite{thr}

\begin{equation}\label{eq:Fsigma}
\begin{aligned}
\frac{d\sigma_{\gamma N\to J/\psi N}}{dt} \Bigg|_{t=t_{thr},W=W_{thr} }
&=\frac{d\sigma_{\gamma N\to J/\psi N}}{dt} \Bigg|_{t=t_{min},W=W_{thr}}
\\
&=\frac{\sigma_{\gamma N\to J/\psi N}( W_{thr} )   }{4 |p_{\gamma} |\cdot  |p_{J/\psi} | }
\end{aligned}
 \end{equation}
where $t_{thr} = m_{J/\psi}^2M_{N}/(m_{J/\psi}+M_{N})$.
It is found that the differential cross section at the threshold energy can be obtained by those three methods. Note that the energy threshold $W=W_{thr}$, as an ideal point, is impossible to obtain the photoproduction data experimentally. Therefore, researching those three methods in Eq.  (\ref{eq:Fsigma}) at a near-threshold energy region is a good window
to study the energy threshold result.
One of the purposes of this paper is to compare the distribution contributions of these three methods.
\begin{table}
\centering
\caption{The values of the  fitted parameters  in the effective Pomeron model   \cite{po2}.}
\label{apsi}
\setlength{\tabcolsep}{5mm}{\begin{tabular}{ccccc}
\hline\hline\noalign{\smallskip}
$A_{J/\psi}$&$\alpha_{0}$&$\alpha^{\prime}\rm(GeV^{-2})$&$b_{0}\rm(GeV^{-2})$\\
\noalign{\smallskip}\hline\noalign{\smallskip}
$0.38\pm0.76$&$0.94$&$0.36$&$0.12$\\
\noalign{\smallskip}
\hline\hline
\end{tabular}}
\end{table}

\begin{table}
\centering
\caption{The values of the  fitted parameters in the parametrized two gluon exchange model  \cite{two2}. }
\label{a0}
\setlength{\tabcolsep}{5mm}{
\begin{tabular}{ccc}
\hline\hline\noalign{\smallskip}
$A_{0}$&$A_{1}$&$A_{2}$ \\
\noalign{\smallskip}\hline\noalign{\smallskip}
$0.228\pm0.045$&$-0.218\pm0.006$&$1.221\pm0.005$ \\
\noalign{\smallskip}
\hline\hline
\end{tabular}}
\end{table}

\section{result and discussion}\label{sec:result}
\begin{figure}[h]
	\centering
	\includegraphics[width=0.495\textwidth]{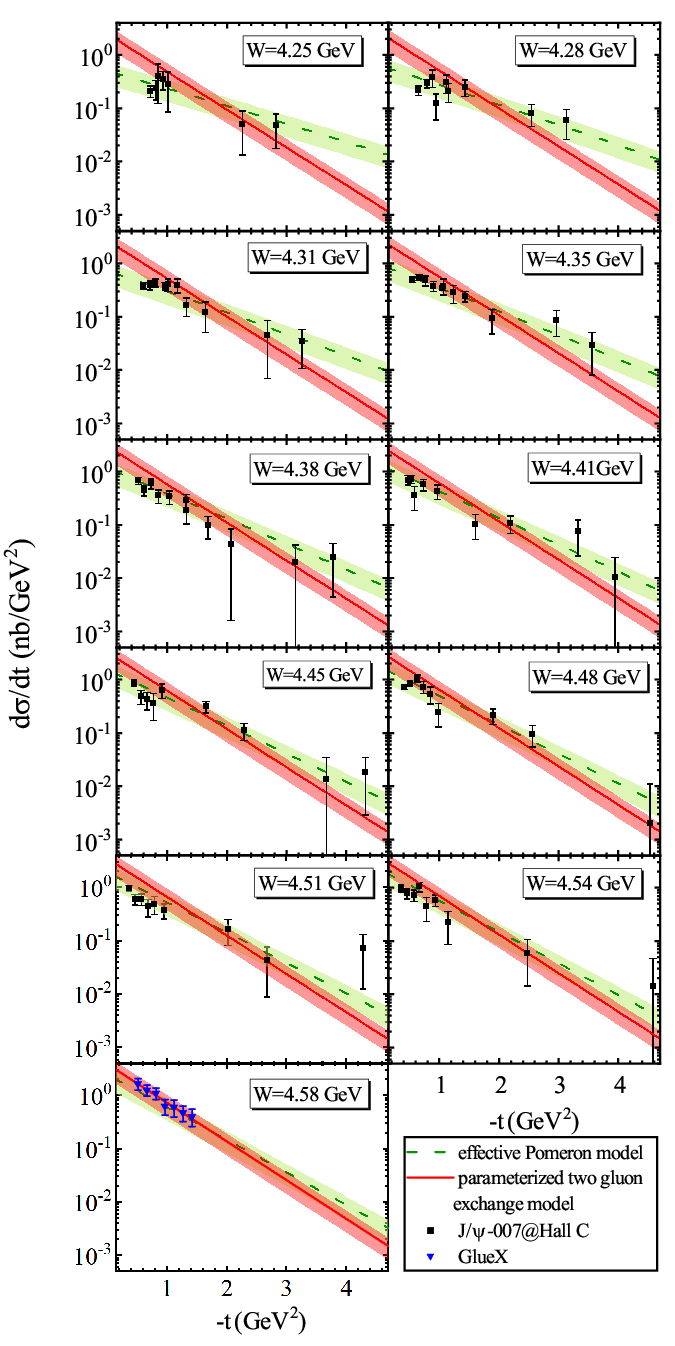}
\caption{The differential cross sections of the $J/\psi$ photoproduction as a function of $-t$ at different c.m. energy. The red solid curve and green dashed line represent the prediction from the parametrized two gluon exchange model and the effective Pomeron model, respectively. The experimental data are from Refs. \cite{new data, glue}.}
  \label{Fig3}
\end{figure}

\begin{figure}[h]
	\centering
	\includegraphics[width=0.48\textwidth]{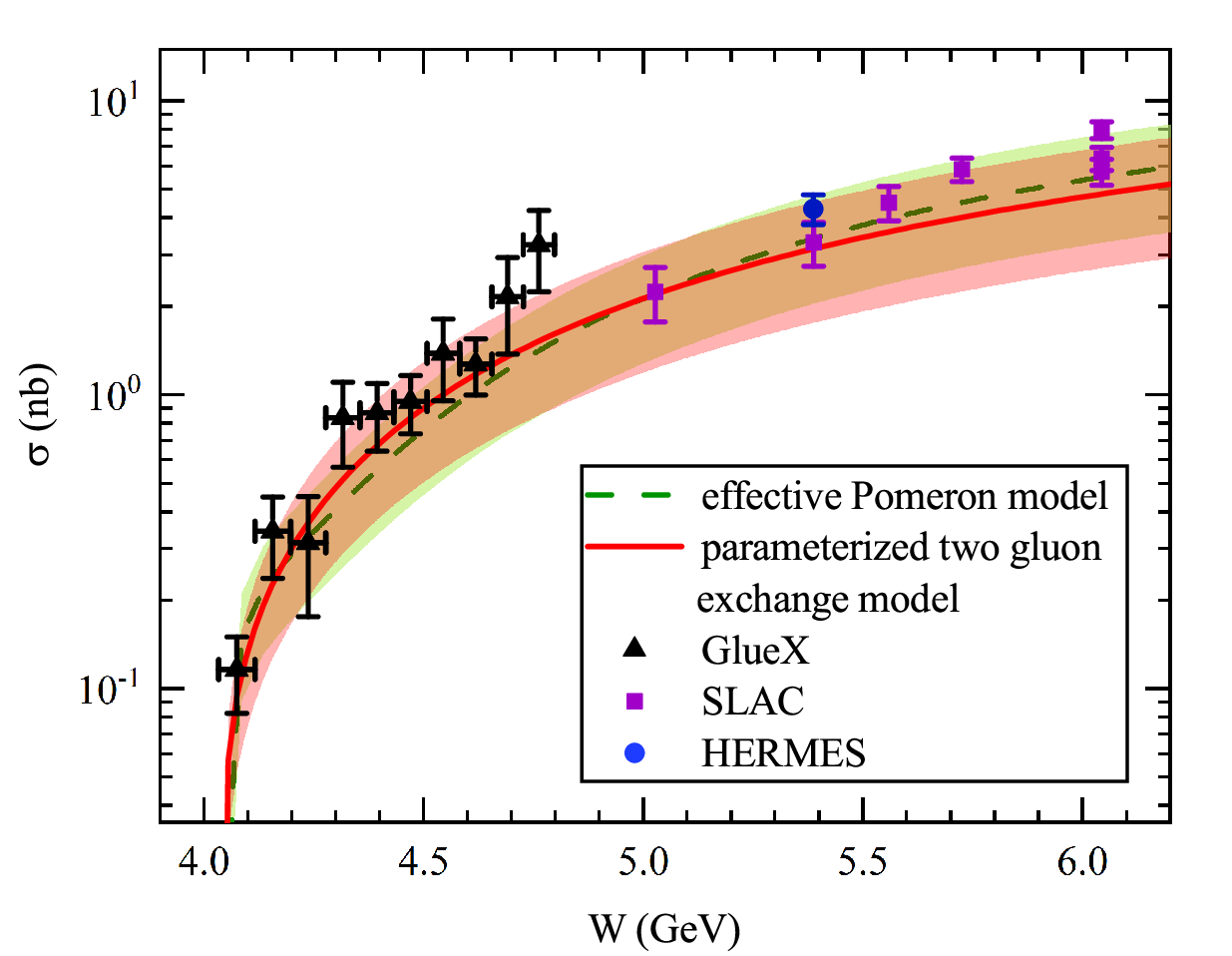}
\caption{The total cross section of the $J/\psi$ photoproduction as a function of $W$.  The curves have the same meaning as in Fig. \ref{Fig3}. The experimental data are from Refs. \cite{glue,slac,hermas}.}
  \label{Fig4}
\end{figure}

In our previous work \cite{two2},
   the free parameters $A_{0},  A_{1},  A_{2}$ of the parametrized two gluon exchange model in 	Eq.  (\ref{eq:do/dt}) were obtained by a global fitting of the total and  differential cross section  data of  $J/\psi$ meson \cite{glue,bd,slac,8,10,11,12,13}, as shown in Table  \ref{a0}. The two models not only explain the $J/\psi$ photoproduction process well, but they also show good consistency  compared with the recent experiment data \cite{new data,glue,bd,slac,8,10,11,12,13,hermas},   which are shown in Figs.  \ref{Fig3} and \ref{Fig4}.

\begin{table}
\centering
\caption{The average value of the QAE contribution $(M_{a}/M_{N})$ extracted  from differential cross section $(t=t_{thr})$. }
\label{avedi}
\setlength{\tabcolsep}{4mm}{
\begin{tabular}{cc}
\hline\hline\noalign{\smallskip}
method&$\ M_{a}/M_{N}(\%)$\\
\noalign{\smallskip}\hline\noalign{\smallskip}
parameterized two gluon exchange model&3.16$\pm$0.63\\
\noalign{\smallskip}\hline\noalign{\smallskip}
effective Pomeron model&3.49$\pm$0.70\\
\noalign{\smallskip}\hline\noalign{\smallskip}
experimental data extraction  &3.50$\pm$0.82\\
\noalign{\smallskip}
\hline\hline
\end{tabular}}
\end{table}

\begin{figure}[h]
	\centering
	\includegraphics[width=0.48\textwidth]{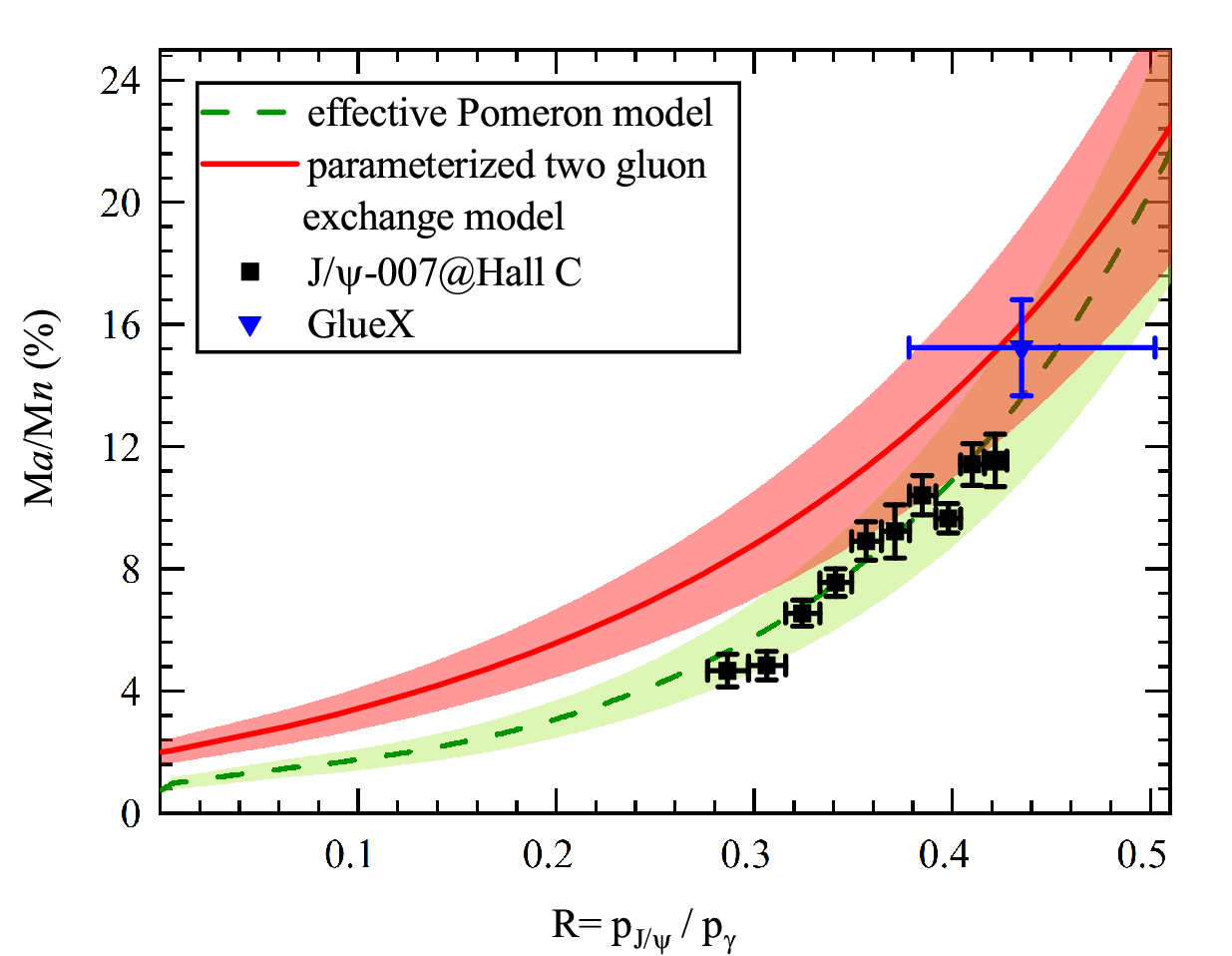}
\caption{The extracted QAE contribution from the differential cross section $(t=t_{\min})$ as a function of $R$. The curves have the same meaning as in Fig. \ref{Fig3}. References of data can be found in \cite{new data, glue}.}
  \label{diffma}
\end{figure}

\begin{figure}[htbp]
	\centering
	\includegraphics[width=0.45\textwidth]{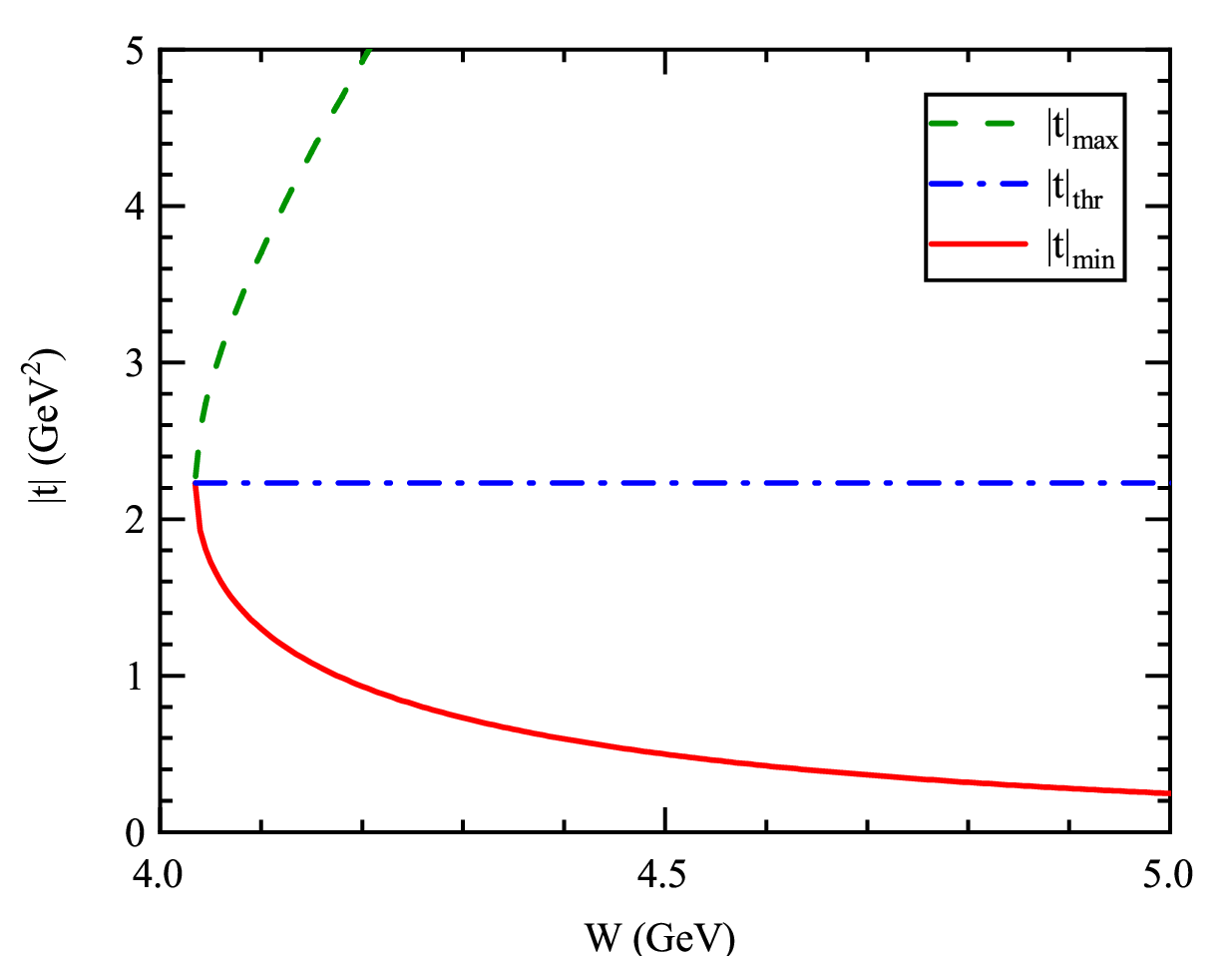}
\caption{The four-momentum transfers $t_{\min},\ t_{\max}\ {\rm and}\ t_{\rm{thr}}$ as a function of $W$.}
  \label{t}
\end{figure}

\begin{table}
\centering
\caption{The average value of the QAE contribution $(M_{a}/M_{N})$ from total cross section.}
\label{averto}
\setlength{\tabcolsep}{4mm}{
\begin{tabular}{cc}\hline\hline\noalign{\smallskip}
method&$M_{a}/M_{N}(\%)$\\
\noalign{\smallskip}\hline\noalign{\smallskip}
parameterized  two gluon exchange model&3.71$\pm$0.74\\
\noalign{\smallskip}\hline\noalign{\smallskip}
effective Pomeron model&3.49$\pm$0.70\\
\noalign{\smallskip}\hline\noalign{\smallskip}
experimental data extraction &4.52$\pm$0.65\\
\noalign{\smallskip}
\hline\hline
\end{tabular}}
\end{table}
First, we consider the case of adopting the differential cross section $d \sigma/dt|_{t=t_{\min}}$ at a near-threshold energy region to study the QAE contribution.
Combined with the predicted differential cross section from theoretical models and the VMD algorithm,
the QAE contribution ($M_a/M_N=(1-b)/4$) as a function of $R$ is obtained as shown in the red-solid and green-dashed curve in Fig. \ref{diffma}.  Besides,  it can be extracted from the GlueX and Hall C experimental data directly \cite{glue,new data}, as shown in the blue triangle and black squares in Fig. \ref{diffma}.
 We note that the QAE contribution to the proton mass is sensitive to energy and varies greatly with energy, which can be seen from both the theoretical models and direct extraction from experimental data.
However, the  energy dependence of the QAE contribution is not the desired conclusion. One can chalk up this energy dependence, in large part, to the  rapid change of $ t_{\min} $ near the threshold, as shown in Fig. \ref{t}.
One find that this effect is less pronounced for the lighter vector mesons  \cite{light}.

\begin{figure}[h]
	\centering
	\includegraphics[width=0.47\textwidth]{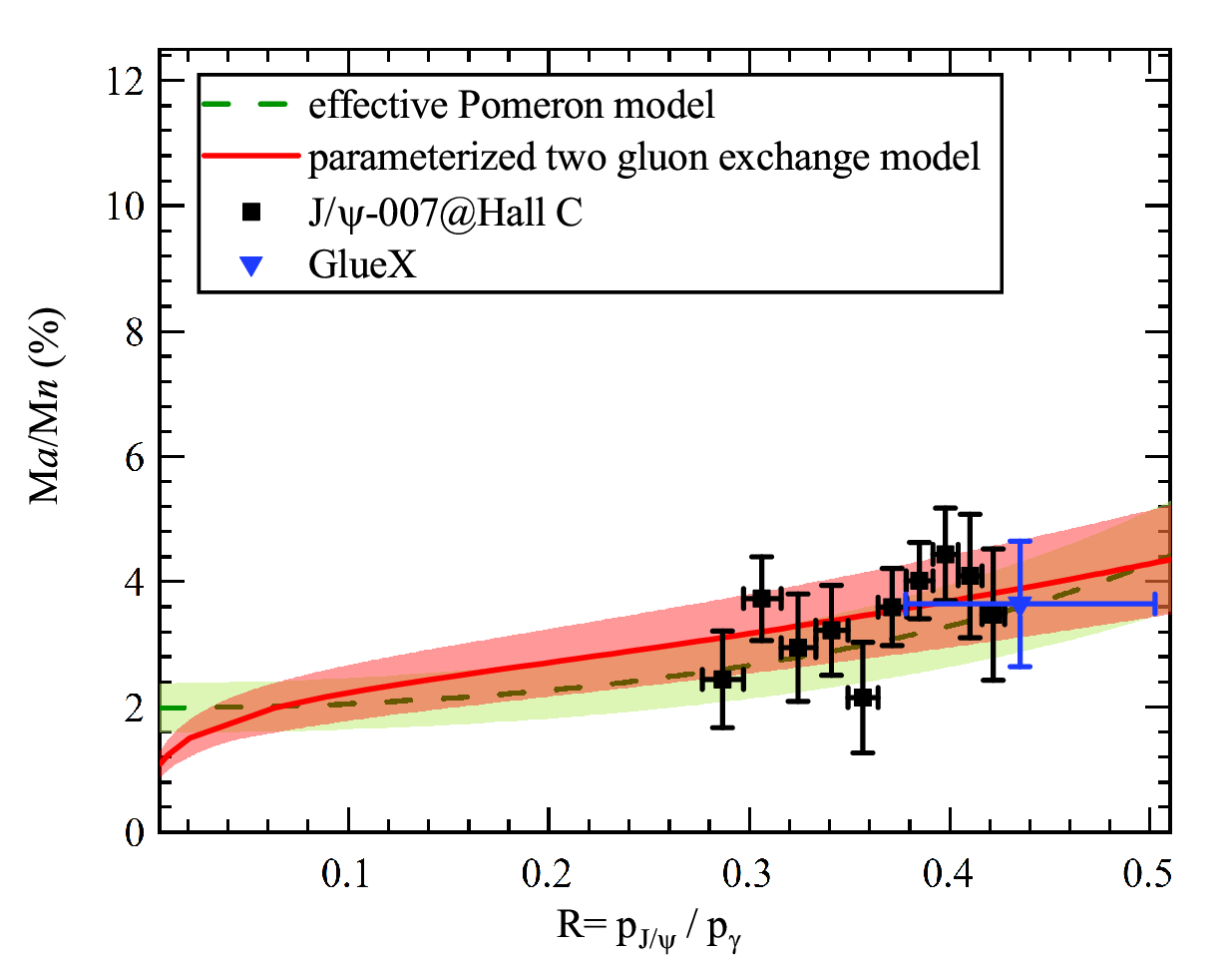}
\caption{The extracted QAE contribution from the differential cross section $(t=t_{\rm{thr}})$ as a function of $R$. The curves have the same meaning as in Fig. \ref{Fig3}. References of data can be found in \cite{new data,glue}.}
  \label{thr}
\end{figure}

Second, the differential cross section $d \sigma/dt|_{t=t_{thr}}$ is  accepted to study the QAE contribution.    Based on theoretical prediction of the cross section data at $t=t_{thr}$, the QAE contribution from the effective Pomeron model and the parametrized two gluon exchange model are calculated, as shown in the red-solid and green-dashed curve in Fig. \ref{thr}, respectively.
The rms value of QAE contribution extracted from the differential cross section at $t=t_{thr}$ within the energy range of $R\in[0,0.5]$  are listed in  Table \ref{avedi}. Note that, $R$  is positively correlated with the c.m. energy, and     $R=0.5$ represents a near-threshold energy that c.m. energy $W=4.79$ GeV.
As depicted in the Fig.  \ref{thr}, the near-threshold stable results to energy   provide us   a good opportunity to study the QAE contribution.

\begin{figure}[h]
	\centering
	\includegraphics[width=0.48\textwidth]{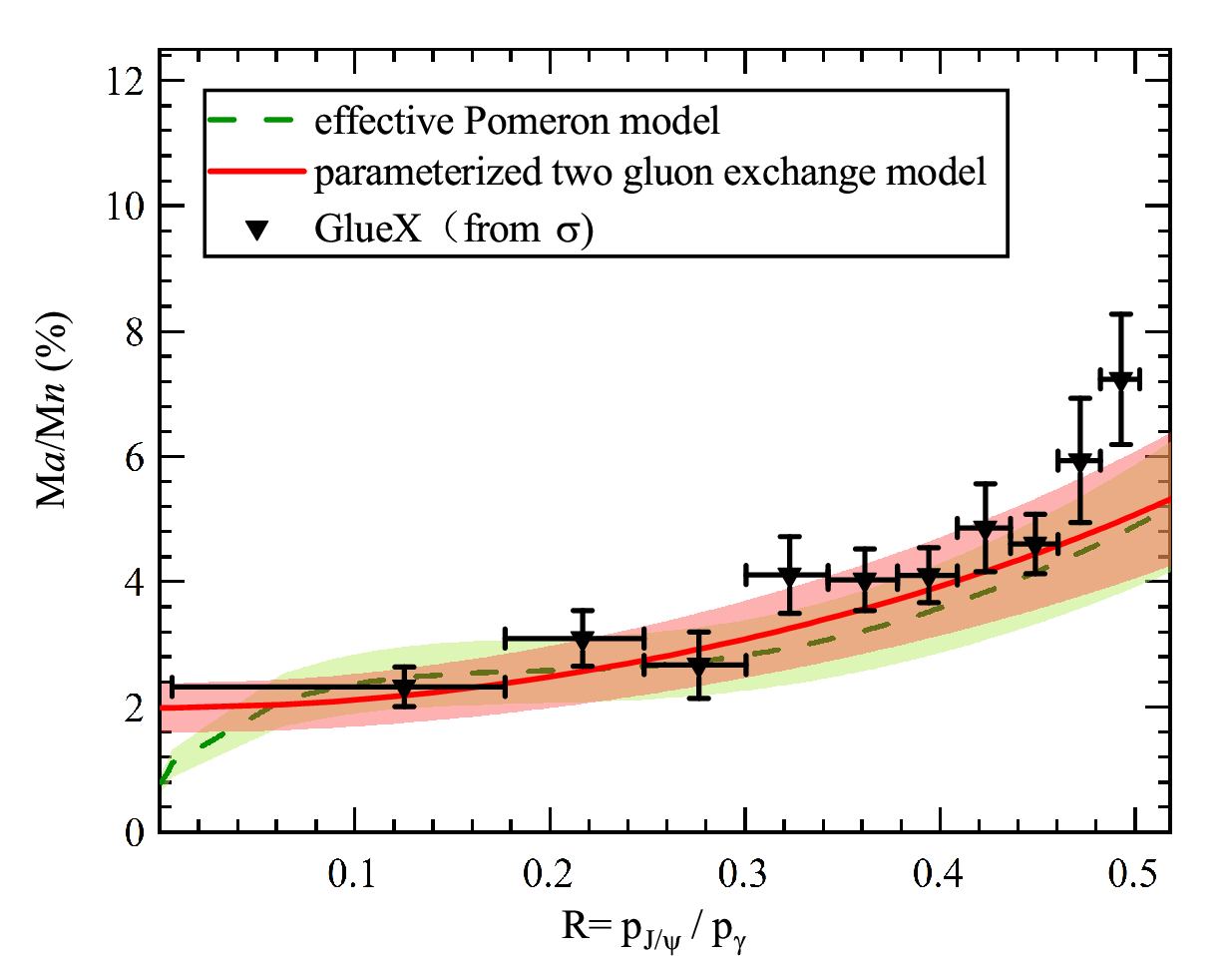}
\caption{The extracted QAE contribution from the total cross section ($t=t_{thr}$) as a function of $R$. The curves have the same meaning as in Fig. \ref{Fig3}. References of data can be found in \cite{glue}.}
  \label{totma}
\end{figure}

The last case uses the total cross section to learn the QAE contribution, which have everything with the second case.
The QAE contribution from theoretical prediction and experimental measurement are listed in Table \ref{averto} and shown in Fig. \ref{totma}.
Similar to the second case,
 the numerical results of the QAE contribution change gently with c.m. energy.
Generally, the last two algorithms provide a robust way to calculate the QAE contribution while
reducing the energy dependence.

In the above study, we find that it is desirable to extract the QAE contribution by calculating the near-threshold total cross section and  the differential cross section when the momentum transfer $t\rightarrow t_{thr}$.
The rms value of the QAE contribution obtained from the total cross section and differential cross section are $(3.66\pm0.72)\%$ and $(3.33\pm0.68)\%$, respectively. Finally, the QAE contribution to the proton  mass  extracted from the $J/\psi$ photoproduction cross section is estimated to be $(3.50\pm0.70)\%$ considering the total and differential cross section results.

\begin{figure}[h]
	\centering
	\includegraphics[width=0.505\textwidth]{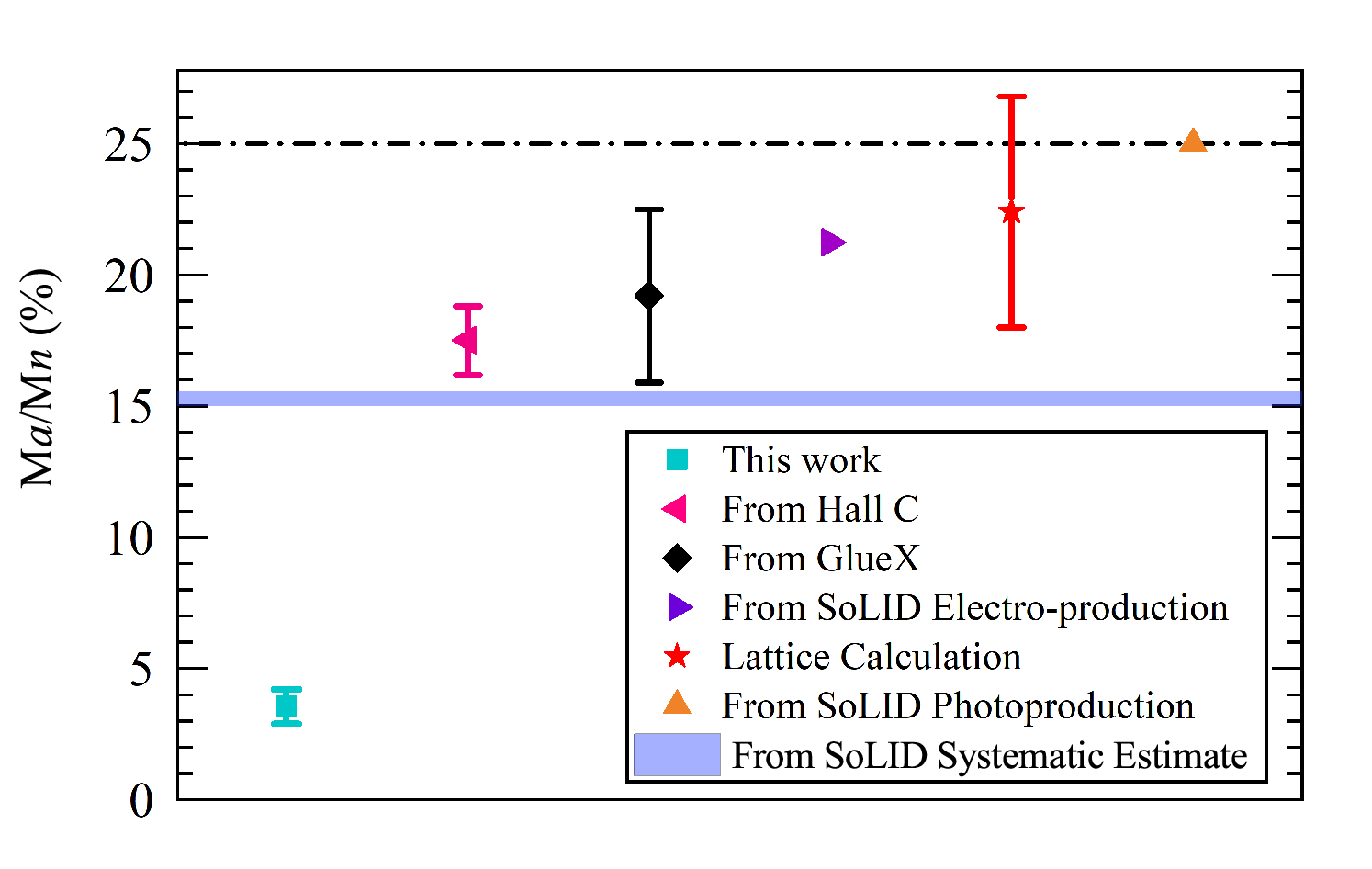}
\caption{Comparison of the QAE contribution from different groups \cite{new data,glue,solid,latice}.}
  \label{compdata}
\end{figure}

By studying the near-threshold differential cross section of vector charmoniums,  some work \cite{new data,wr,solid} extract the QAE contribution using the $J/\psi$ photoproduction differential cross section data at $t=0$  under the framework of VMD model, as shown in Fig. \ref{compdata}.  This method may have a small effect using the light vector meson photoproduction to study the QAE contribution, but results in a significant deviation using the heavy quarkonium photoproduction.  In this work,  we avoid extracting the QAE contribution from the differential cross section  at nonphysical points, momentum transfer  $t = 0$ mentioned in Refs. \cite{wr,new data}.
The feasible approaches  to study the QAE  contribution using the heavy meson photoproduction have been found,  as described in the second and third method above.

\section{SUMMARY}\label{sec:sum}

Using the theoretical predictions of the $J/\psi$ photoproduction cross section within  the effective Pomeron model and the parametrized two gluon exchange model,  a systematic analysis of the QAE  contribution to the proton mass is carried out under the framework of the VMD model.
The results show that these two models can explain  the recent Hall C and GlueX measurements of $J/\psi$ photoproduction well.
 Based on the predicted cross section given by the two models, the distribution of the QAE contribution with the energy is extracted for the first time.
After comparing the results from three methods in Eq. (\ref{eq:Fsigma}), it is found that the QAE contribution extracted from the differential cross section  $d \sigma/dt|_{t=t_{\min}}$ has a strong energy dependence.  One can chalk up this energy dependence to the  rapid change of $t_{\min}$ near the threshold.
Meanwhile,
the  near-threshold total cross section and differential cross section $d \sigma/dt|_{t=t_{thr}}$ provide a feasible  way to calculate the QAE contribution while reducing the energy dependence.
This gives us a better window to study the energy threshold result through the near-threshold vector meson production data.
Finally, the average value of the QAE contribution is estimated to be (3.50$\pm$0.70)$\%$, which suggests that the QAE contribution to the proton mass is small.
 Accordingly, we compared this result with those of other groups and explored the reasons for the differences.
Our work is underestimated with the results in Refs. \cite{new data,wr,solid}, and comparable with the conclusion in Ref. \cite{am}. Moreover, one note that in Ref. \cite{hy}, the contribution of the QED EMT trace anomaly to the hydrogen atom mass was calculated, and the obtained values are small and close to our results. These situations indicate that the magnitude of the QAE contribution is still uncertain, and more experimental and theoretical studies are still needed.

\section{}

X.-Y. Wang appreciated the discussion with Dr. Daniel Winney about the efficient Pomeron model developed by the JPAC collaboration. This work is supported by the National Natural Science Foundation of China under Grants No. 12065014 and No. 12047501, and by the Natural Science Foundation of Gansu province under Grant No. 22JR5RA266. We acknowledge the West Light Foundation of The Chinese Academy of Sciences, Grant No. 21JR7RA201.

\end{document}